\documentstyle[prd,aps,twocolumn]{revtex}
\begin{document}
\title{Naked singularities  and Seifert's conjecture\thanks 
               {This paper is dedicated to Professor P. C. Vaidya on
                the occasion of his 80th birthday.}}
\author{K. S. Virbhadra\thanks{Email address : shwetket@maths.uct.ac.za}}
\address{Department of Applied Mathematics,
                    University of Cape Town,
                    Rondebosch 7701,
                    South Africa                
                    }
\maketitle
\begin{abstract} 
It is shown that for a general nonstatic spherically symmetric metric of 
the Kerr-Schild class several energy-momentum complexes  give the same energy 
distribution as  in the Penrose prescription, obtained by Tod. This result is
useful for investigating the Seifert conjecture for naked singularities. The
naked singularity forming in the Vaidya null dust collapse supports the
Seifert conjecture.  Further, an example and a counterexample to this 
conjecture are presented in the Einstein massless scalar theory.
\end{abstract}

In a series of seminal papers it was established that under a wide variety
of circumstances a spacetime singularity is inevitable in a physically 
realistic complete gravitational collapse\cite{HE73}. However, these
studies were silent on a very important question: whether or not a spacetime
singularity arising therefrom will be visible to observers. To this end
about three decades ago, Penrose\cite{Pen69} in his review on gravitational
collapse asked  `` Does there exist a cosmic censor who forbids the occurrence
of naked singularities, clothing each one in an absolute event horizon?''
Though there is no precise statement of a {\em cosmic censorship hypothesis} 
(CCH), roughly speaking it states that naked singularities do not occur in 
generic realistic gravitational collapse. There are two versions of the CCH:
the hypothesis that, generically, singularities forming due to gravitational
collapse are hidden inside black holes is known as the {\em weak cosmic
censorship hypothesis}, whereas the {\em strong cosmic censorship hypothesis}
states that, generically, timelike singularities never occur (see \cite{Wal97}
and references therein). No proof for any version of the CCH is
known  and it is considered  the most important unsolved problem in classical
general  relativity \cite{Haw79,Isr84,Pen98}. The subject of singularities  
fascinated many  researchers (see 
\cite{HE73,Pen69,Wal97,Haw79,Isr84,Pen98,Sing,NS,Jos97,Sin98,Des98,Iguetal98,Viretal98,Pen73,Sei79,Chi94}
and references therein).

Since the CCH was proposed, numerous counterexamples to this hypothesis
have been found in the literature (see \cite{NS,Jos97,Sin98,Des98} and 
references therein). Iguchi {\it et al.} \cite{Iguetal98} examined the 
stability of nakedness of singularity in the Lemaitre-Tolman-Bondi collapse 
against the odd-parity modes of nonspherical linear perturbations for the 
metric and reported that the 
perturbations do not diverge but are well defined even in the neighbourhood
of a central naked singularity. Much  work has been done on black holes
as well as on naked singularities, but it was not known how black holes
could be differentiated observationally from naked singularities (if these
indeed exist). Recently it has been shown that the gravitational lensing
could be possibly helpful for this purpose\cite{Viretal98}. 
As the known counterexamples to CCH are of special geometric nature, it is  
still believed by most that the conjecture correctly characterizes  realistic 
gravitational collapse. Recently, Wald\cite{Wal97}
reviewed the status of the weak cosmic censorship and expressed his view
that naked singularities cannot arise generically.

The idea of naked singularities has been disliked by many physicists, as 
their existence is thought to give serious problems.  For instance, there
can be production of matter and/or radiation out of extremely high 
gravitational fields and the mechanism for that is not known
and there can be other causal influences from the infinitely compressed
material. When the first
few counterexamples to the CCH were obtained it was clear that the CCH 
could not be proved in generality. Penrose\cite{Pen73} imposed a condition
that a naked singularity must be proved to be stable against perturbations
of initial conditions as well as equations of state.
Seifert\cite{Sei79} conjectured
that any singularity that occurs if a finite nonzero amount of matter
tends to collapse into one point is hidden and naked singularities occur
only if one has singularities along lines or surfaces or if the central
singularities are carefully arranged that they contain only zero mass.

There have been some discussions on  ``mass'' of naked singularities
\cite{Isr84,Chi94}, but not many  studies have been done.
The investigation of this subject is difficult as there
is no agreed and precise definition of  local or quasilocal mass  in
general
relativity and it has been a ``recalcitrant'' problem since the outset
of this theory. Bergqvist\cite{Ber92} performed calculations with
several different  definitions of  mass (their uses are not restricted
to ``Cartesian coordinates'') and reported that not  any two of them give 
the same result for the Reissner-Nordstr\"om  and Kerr spacetimes. On the
other hand it is also known that several  {\em energy-momentum complexes}
(their uses are restricted to ``Cartesian coordinates'')  give the same and
reasonable results for some well known spacetimes \cite{KSV,VirCham,ACV96}.
These are encouraging results and we will use  some of them to investigate
the Seifert conjecture.   We use geometrized 
units and follow the convention that Latin (Greek) indices run from
$0\ldots3$ ($1\ldots3$). The comma and semicolon, respectively, stand for 
the partial and covariant derivatives.

The Einstein energy-momentum complex is \cite{Mol5861}

\begin{equation}
\Theta_i{}^{k} = \frac{1}{16 \pi} H^{\ kl}_{i \ \ ,l} ,
\label{eq1}
\end{equation}
where
\begin{equation}
H_i^{\ kl}\  =\ - H_i^{\ lk}\ =\  \frac{g_{in}}{\sqrt{-g}}
         \left[-g \left( g^{kn} g^{lm} - g^{ln} g^{km}\right)\right]_{,m} \ .
\label{eq2}
\end{equation}
$\Theta_0^{\ 0}$ and $\Theta_{\alpha}^{\ 0}$ stand for the energy and momentum
density components, respectively.\footnote{Though the energy-momentum complex
obtained by Tolman differs in form from the Einstein energy-momentum complex, 
both are equivalent in import\cite{Tol30}. The present author was earlier 
unaware of this \cite{KSV,VirCham,ACV96}.} The energy and momentum components 
are given by
\begin{equation}
P_i \ = \ \int \int \int \Theta_i^{\ 0} dx^1  dx^2 dx^3 .
\label{eq3}
\end{equation}
Applying Gauss's theorem one obtains 
\begin{equation}
P_i\ =\ \frac{1}{16 \pi} \ \int\int\ H_i^{\ 0 \alpha} \ n_{\alpha}\ dS \text{,}
\label{eq4}
\end{equation}
where $n_{\alpha}$ is the outward unit normal vector over the infinitesimal 
surface element $dS$  and $P_0$ and $P_{\alpha}$ stand for the energy and 
momentum components, respectively.

The symmetric energy-momentum complex of Landau and Lifshitz is\cite{LL87}
 \begin{equation}
L^{ik}=  \frac{1}{16 \pi} {{\lambda}^{iklm}}_{,lm},
\label{eq5}
 \end{equation}
where
 \begin{equation}
{\lambda}^{iklm}=-g \left(g^{ik} g^{lm}-g^{il} g^{km}\right).
\label{eq6}
 \end{equation}
$L^{00}$ and $ L^{\alpha 0}$  are the energy and energy current
(momentum) density components. 
The energy and momentum  are given by
 \begin{equation}
P^i=\int\!\!\!\int\!\!\!\int{L^{i0}\,dx^1\,dx^2\,dx^3} .
\label{eq7}
 \end{equation}
Using Gauss's theorem, the energy and momentum components are
 \begin{equation}
P^i=\frac{1}{16 \pi} 
   \int\!\!\!\int{ {{\lambda}^{i0\alpha m}}_{,m} \  n_{\alpha}\ dS} .
\label{eq8}
 \end{equation}

The symmetric energy-momentum complex  of Papapetrou is\cite{Pap48}
 \begin{equation}
\Sigma^{ik}=\frac{1}{16 \pi}{N^{iklm}}_{,lm},
\label{eq9}
 \end{equation}
where
 \begin{equation}
N^{iklm}=\sqrt{-g} \left(g^{ik} \eta^{lm} - g^{il} \eta^{km}
+ g^{lm} \eta^{ik} - g^{lk} \eta^{im}\right),
\label{eq10}
 \end{equation}
with
 \begin{equation}
\eta^{ik}= {\rm diag}(1,-1,-1,-1).
\label{eq11}
 \end{equation}

$\Sigma^{00}$ and $\Sigma^{\alpha 0}$  are the energy  and energy
current (momentum) density components. 
The energy and momentum components are given by
 \begin{equation}
P^i=\int\!\!\!\int\!\!\!\int{\Sigma^{i0}\,dx^1\,dx^2\,dx^3} .
\label{eq12}
 \end{equation}
For the time-independent metrics, one has
 \begin{equation}
P^i=\frac{1}{16 \pi} 
     \int\!\!\!\int{{N^{i0\alpha\beta}}_{,\beta} \  n_{\alpha}\  dS} .
\label{eq13}
 \end{equation}

The symmetric energy-momentum complex of Weinberg is\cite{Wei72}
 \begin{equation}
W^{ik}= \frac{1}{16 \pi} {D^{lik}}_{,l},
\label{eq14}
 \end{equation}
where
 \begin{eqnarray}
D^{lik}= \frac{\partial h^a_{\ a}}{\partial x_l}\eta^{ik}
         &-&  \frac{\partial h^a_{\ a}}{\partial x_i}\eta^{lk} 
         - \frac{\partial h^{al}}{\partial x^a}\eta^{ik}
         + \frac{\partial h^{ai}}{\partial x^a}\eta^{lk}  
         + \frac{\partial h^{lk}}{\partial x_i}  \nonumber\\
         &-& \frac{\partial h^{ik}}{\partial x_l}
\label{eq15}
 \end{eqnarray}
and
 \begin{equation}
h_{ik}=g_{ik} - \eta_{ik}.
\label{eq16}
 \end{equation}
$\eta_{ik}$ is the Minkowski metric. Indices on $h_{ik}$ or $\partial/
\partial x_i$ are raised or lowered with help of $\eta$'s.
It is clear that
 \begin{equation}
D^{lik}=- D^{ilk}.
\label{eq17}
 \end{equation}

$W^{00}$ and $W^{\alpha 0}$  are the energy and energy current
(momentum) density components. 
The energy and momentum components  are given by
 \begin{equation}
P^i=\int\!\!\!\int\!\!\!\int{ W^{i0}\, dx^1 \, dx^2 \, dx^3} .
\label{eq18}
 \end{equation}
Using Gauss's theorem, one has
 \begin{equation}
P^i=\frac{1}{16 \pi} \int\!\!\!\int{ D^{\alpha 0 i} \  n_{\alpha}\   dS} .
\label{eq19}
 \end{equation}

Though the uses of the energy-momentum complexes are restricted to ``Cartesian
coordinates'' (i.e., these give meaningful results  in these coordinates), 
these satisfy the {\em local conservation laws}
($\Theta_i{}^{k}{}_{,k} = 0,\  L^{ik}_{\ ,k} = 0,\ \Sigma^{ik}_{\ ,k} = 0,
\ W^{ik}_{\ ,k} = 0$) in all system of coordinates.

We first discuss some of  our earlier results in brief. Then  we  use them
to show that the energy-momentum complexes of Einstein,  Landau and
Lifshitz, Papapetrou and Weinberg, and the Penrose quasilocal definition
 give the same result for a general nonstatic
spherically symmetric metric  of the Kerr-Schild class.
The Kerr-Schild class spacetimes are given by metrics $g_{ik}$ of the form
\begin{equation}
g_{ik} = \eta_{ik} - H l_i l_k ,
\label{eq20}
\end{equation}
where $\eta_{ik} = {\rm diag}(1,-1,-1,-1)$ is the Minkowski metric.
$H$ is the scalar field and $l_i$ is a null, geodesic and shear free
vector field in the Minkowski spacetime, which are respectively
expressed as 
\begin{eqnarray}
\eta^{ab} l_a l_b &=& 0 , \nonumber\\
\eta^{ab} l_{i,a} l_b &=& 0 , \nonumber \\
\left(l_{a,b} + l_{b,a}\right) {l^a}_{,c} \  \eta^{bc}
 - \left({l^a}_{,a}\right)^2 &=& 0. 
\label{eq21}
\end{eqnarray}

An interesting feature of the Kerr-Schild class metric $g_{ik}$ in 
Eq. $(\ref{eq20})$
is that the vector field $l_i$ remains null, geodesic and shear free
with the metric $g_{ik}$. Equations  $(\ref{eq21})$ lead to 
\begin{eqnarray}
g^{ab} l_a l_b &=& 0 , \nonumber\\
g^{ab} l_{i;a} l_b &=& 0 , \nonumber \\
\left(l_{a;b} + l_{b;a}\right) {l^a}_{;c} \  g^{bc}
 - \left({l^a}_{;a}\right)^2 &=& 0. 
\label{eq22}
\end{eqnarray}

There are several well-known spacetimes of the Kerr-Schild class,
for instance, Schwarzschild, Reissner-Nordstr\"{o}m, Kerr, Kerr-Newman,
Vaidya, Dybney {\it et al.}, Kinnersley, Bonnor-Vaidya, and  Vaidya-Patel (for
references see in \cite{Vai74}).

It is known that the energy-momentum complexes of Einstein  $\Theta_i{}^{k}$,
 Landau and Lifshitz $L^{ik}$, Papapetrou 
$\Sigma^{ik}$ and Weinberg $W^{ik}$ ``coincide'' for any Kerr-Schild class
metric\cite{ACV96}. 

These energy-momentum complexes for any Kerr-Schild class metric are given by\cite{ACV96}
 \begin{eqnarray}
\Theta_i{}^{k} &=&   \eta_{ij} L^{jk},\label{eq23} \\
L^{ik} &=& \Sigma^{ik} = W^{ik} =
\frac{1}{16\pi}\Lambda^{iklm}{}_{,lm},
\label{eq24}
\end{eqnarray}
where
 \begin{eqnarray}
\Lambda^{ikpq}&\equiv&H\left(
\eta^{ik}l^pl^q+\eta^{pq}l^il^k-\eta^{ip}l^kl^q-\eta^{kq}l^il^p
\right).
\label{eq25}
\end{eqnarray}

To obtain the above result  for the Landau and Lifshitz, Papapetrou and
Weinberg complexes in terms of the scalar function $H$ and the vector $l_i$,
only null condition of Eqs.  $(\ref{eq21})$ was used while for the Einstein 
complex the null as well as geodesic conditions  were used. The shear-free
condition was not required to obtain Eqs.  $(\ref{eq23})$-$(\ref{eq25})$.
Thus, these energy-momentum complexes ``coincide'' for a class of solutions
more general than the Kerr-Schild class.
The energy and momentum components are
 \begin{equation}
P^i=\frac{1}{16 \pi} 
\int\!\!\!\int{ {{\Lambda}^{i0\alpha m}}_{,m} \ n_{\alpha}\,dS} .
\label{eq26}
 \end{equation}

 The energy-momentum complexes of Landau 
and Lifshitz, Papapetrou and Weinberg are symmetric in their indices and 
therefore have been used to define angular momentum; the spatial components
of angular momentum are (though we do not use  this in this  paper  we 
give here  for  completeness)
 \begin{equation}
J^{\alpha \beta}=\frac{1}{16 \pi} \int\!\!\!\int{
\left(x^{\alpha}{{\Lambda}^{ \beta0 \sigma m}}_{,m}
-x^{\beta}{{\Lambda}^{ \alpha0 \sigma m}}_{,m}
+\Lambda^{ \alpha 0 \sigma\beta}
\right)n_{\sigma}\,dS}.
\label{eq27}
\end{equation}

Now we consider a general nonstatic spherically symmetric spacetime of the 
Kerr-Schild class given by the line element
\begin{equation}
ds^2 = B\left(u,r\right) du^2 - 2 du dr 
     - r^2 \left(d\theta^2+\sin^2\theta d\phi^2\right) .
\label{eq28}
\end{equation}

Transforming the above line element in $t,x,y,z$ coordinates according to
\begin{eqnarray}
u &=& t+r, \nonumber\\
x\ &=& \ r\ \sin\theta\ \cos\phi, \nonumber\\
y\ &=& \ r\ \sin\theta \ \sin\phi, \nonumber\\
z\ &=& \ r\ \cos\theta,
\label{eq29}
\end{eqnarray}
one gets
\begin{eqnarray}
ds^2 &=& dt^2 - dx^2 -  dy^2 - dz^2 - 
        \left(1 - B\left(t,x,y,z\right)\right)    \nonumber\\
&& \times \left[dt + \frac{x dx + y dy +z dz}{r}\right]^2.
\label{eq30}
\end{eqnarray}

This is obviously a Kerr-Schild class metric with $H = 1 - B$ and $l_i
= \left(1, x/r, y/r, z/r \right)$. Using these  in Eqs.  $(\ref{eq25})$
 and $(\ref{eq26})$ one gets the expression  for energy
\begin{equation}
E = \frac{r}{2} \left(1-B\left(u\right)\right).
\label{eq31}
\end{equation}

Tod\cite{Tod}, using the Penrose quasilocal mass definition, got the same 
result. This is indeed an encouraging result. The Penrose mass is called 
quasilocal because it is obtained over 
two-surface and  is not found by an integral over a spanning three-surface 
of a local density. The above expression can be used to test the Seifert 
conjecture for the naked singularities arising due to a general spherical 
collapse
described  by  Eq. $(\ref{eq28})$ and satisfying the {\em weak energy condition}.
These investigations could give conditions on $B(u)$ for  the Seifert 
conjecture to be true and to be false. It could be possible that the Seifert
conjecture is true in all these cases.
The Vaidya null dust collapse is extensively studied (see \cite{Jos97} and
references therein). $B = 1 - 2 M(u)/r$ in Eq. $(\ref{eq28})$ gives
the Vaidya null dust collapse solution. For this,  Eq. $(\ref{eq31})$ gives
$E = M(u)$ (see also \cite{VirCham}). It is known that for $ M = 
\lambda u$,  a naked singularity occurs at $r=0, u=0$ for $\lambda \le 1/16$.
Physically, $dM/du$ represents the power (energy flowing per unit time)
imploding through a two-sphere of radius $r$ and it  must be non-negative for 
the {\em weak energy condition} to
be satisfied. Using Eq.  $(\ref{eq31})$ one finds that the naked singularity in
the Vaidya null dust collapse is massless. This  supports the Seifert
conjecture. For the Bonnor-Vaidya spacetime $B = 1 - 2 M(u)/r
+{Q\left(u\right)}^2/r^2$, where $Q(u)$ is the charge parameter.
This represents charged null dust collapse. 
The result in  Eq.  $(\ref{eq31})$ can be useful for investigations of the
Seifert conjecture in the Bonnor-Vaidya collapse.

Now we wish to investigate the Seifert conjecture in the Einstein massless
scalar (EMS) theory, given by equations
\begin{equation}
R_{ij}\ -\ \frac{1}{2}\ R \ g_{ij}\ =\ 8 \pi \ T_{ij}\ 
\label{eq32}
\end{equation}
and
\begin{equation}
\Phi_{,i}^{\ ;i}\ =\ 0 .
\label{eq33}
\end{equation}

 $R_{ij}$ is the Ricci tensor and $R$ is the Ricci scalar. 
$\Phi$ stands for the massless scalar field.
 $T_{ij}$, the energy-momentum tensor of the massless scalar field, is
given by
\begin{equation}
T_{ij}\ =\ \Phi_{,i}\ \Phi_{,j}\ -\ \frac{1}{2}\ g_{ij}\ g^{ab}\ \Phi_{,a}\
          \Phi_{,b} .
\label{eq34}
\end{equation}

Equation $(\ref{eq32})$ with Eq. $(\ref{eq34})$ can be expressed as
\begin{equation}
R_{ij}\ =\ 8 \pi \ \Phi_{,i}\ \Phi_{,j} .
\label{eq35}
\end{equation}

 It is known that the most general static spherically
symmetric solution to the EMS equations (with the cosmological constant
$\Lambda = 0$) is asymptotically flat and until recently it was known that
this is the well-known Wyman solution\cite{Rob93}. Recently, it has been
shown that the Janis-Newman-Winicour (JNW) solution (which was obtained about
thirteen years before the Wyman solution) is the same as the Wyman 
solution\cite{Vir97}.  This solution is characterized by two  constant
parameters, the mass $M$ and  the ``scalar charge'' $q$, and  is given by the
line element (see in \cite{Vir97})
\begin{eqnarray}
ds^2 &=& \left(1-\frac{b}{r}\right)^{\gamma} dt^2 
      - \left(1-\frac{b}{r}\right)^{-\gamma} dr^2 \nonumber\\
      &-& \left(1-\frac{b}{r}\right)^{1-\gamma}  
           r^2 \left(d\theta^2  +\sin^2\theta \  d\phi^2\right)
\label{eq36}
\end{eqnarray}
and the scalar field
\begin{equation}
\Phi = \frac{q}{b\sqrt{4\pi}} \ln\left(1-\frac{b}{r}\right),
\label{eq37}
\end{equation}

where
\begin{eqnarray}
\gamma &=& \frac{2M}{b}, \nonumber\\
b &=& 2 \sqrt{M^2+q^2}.
\label{eq38}
\end{eqnarray}

For the ``scalar charge'' zero this solution reduces to the Schwarzschild 
solution. $r = b$ is a globally naked strong curvature singularity
(see \cite{Viretal97} with \cite{Vir97}). It is of interest to investigate
whether or not this naked singularity is massless. Obviously Eq. $(\ref{eq31})$
cannot be used for this spacetime. In the following we obtain the energy 
expression for the most  general nonstatic spherically symmetric metric
described by the line  element
\begin{eqnarray}
ds^2\ &=& B\left(r,t\right) dt^2 - A\left(r,t\right) dr^2 
       - 2 F\left(r,t\right) dt dr  \nonumber\\
  &-& D\left(r,t\right) r^2 \left(d\theta^2 + \sin^2\theta d\phi^2\right).
\label{eq39}
\end{eqnarray}

We transform the  line element $(\ref{eq39})$ to ``Cartesian coordinates''
$t,x,y,z$ 
(according to $x =  r \sin\theta \cos\phi, y = r \sin\theta  \sin\phi, 
z =  r \cos\theta$) and then  calculate energy distribution associated with
this using definitions of Einstein, Landau and Lifshitz, and Weinberg; these
are respectively given by
\begin{eqnarray}
E_E &=& \ \frac{r\left[B\left(A-D-D'r\right)-F\left(r\dot{D}-F\right)\right]}
           {2 \sqrt{AB+F^2}} , \label{eq40} \\
E_{LL} &=& \ \frac{ r D \left(A - D - r D' \right) }
                {2} , \label{eq41} \\
E_W &=& \    \frac{ r  \left(A - D - r D' \right) }
                {2} .
\label{eq42}
\end{eqnarray}
The prime and dot denote for the partial derivative with respect to the
coordinates $r$ and $t$, respectively. 
These energy-momentum complexes have  an advantage that one can easily apply
Gauss's theorem to get energy contained inside a two-surface for any spacetime.
The same is possible with the Papapetrou energy-momentum complex for
the Kerr-Schild class metrics, but not for arbitrary spacetimes. However,
while using the Papapetrou energy-momentum complex one can apply Gauss's 
theorem for any time-independent metric [see Eq.  $(\ref{eq13})$]. Thus, for static
and spherically symmetric  spacetimes ($F = 0$ and $A, B$ and $D$ depending
only on radial coordinate) one gets
\begin{eqnarray}
E_P &=&  \frac{r}{8 (AB)^{3/2}} 
    \left[4 A B^2 (A-D) 
  +  r(A^2 B' D -  2 A^2 B D' \right. \nonumber\\ 
  &-& \left.  A A' B D 
   -  2  A B^2 D' -  A B B' D +  A' B^2 D) \right] .
\label{eq43}
\end{eqnarray}

In a series of papers there are encouraging results (see in 
\cite{KSV,VirCham,ACV96}). Several energy-momentum complexes give the
same results for the Einstein-Rosen as well as any metric of the Kerr-Schild
class (the Kerr-Schild class has several well-known solutions).
As mentioned earlier these energy-momentum complexes ``coincide'' for a more
general class than the Kerr-Schild class. However,
it is obvious from Eqs.  $(\ref{eq40})$- $(\ref{eq43})$  that the energy-
momentum complexes disagree for the most general nonstatic spherically
symmetric metric. While obtaining
the energy distribution (given by equation $(\ref{eq31})$) for the spacetime
described by the line element $(\ref{eq28})$ we used Kerr-Schild Cartesian
coordinates; however, for the case of the most general nonstatic spherically
symmetric metric we used what we call ``Schwarzschild Cartesian coordinates.''
The line element $(\ref{eq28})$ is a special case of the line element 
$(\ref{eq39})$. The Einstein energy-momentum complex gives the same result
in both the cases (consider Eq. $(\ref{eq31})$ as a special case of  Eq.
$(\ref{eq40})$) which is the same as Tod found using the Penrose definition.
However, other definitions disagree with their own results obtained in 
Kerr-Schild Cartesian  and Schwarzschild Cartesian coordinates
(as Eqs.  $(\ref{eq41})$, $(\ref{eq42})$ and $(\ref{eq43})$ do not yield Eq. 
$(\ref{eq31})$ as a special case). For a simple case of the Schwarzschild
metric, $E_{LL} = E_P = E_W = M$ when calculations are performed in the
 Kerr-Schild
Cartesian coordinates; however, in Schwarzschild Cartesian coordinates one
has $E_{LL} = E_W = M \left(1-2M/r\right)^{-1}, 
E_P = M\left(r^2-2Mr+2M^2\right)/ \left(r-2M\right)^2$. It is not clear
why different energy-momentum complexes ``coincide'' in the Kerr-Schild 
Cartesian coordinates, but not in the Schwarzschild Cartesian coordinates. 
The Einstein- Rosen metric is not  of the Kerr-Schild class and therefore 
calculations were  not done in Kerr-Schild Cartesian coordinates; however, different 
energy-momentum complexes gave the same and reasonable result.  Therefore, the 
fact that different energy-momentum complexes give the same result for some 
spacetimes  is not restricted  to the use of the  Kerr-Schild Cartesian 
coordinates. It is known that the quasilocal mass definitions also have
some problems (see in \cite{Ber92,BT94}); for instance, it has not
been possible to obtain the Penrose quasilocal mass for the Kerr metric.

The present  investigations indicate that the Einstein energy-momentum complex
is  better  than other energy-momentum complexes. Therefore, the result in Eq.
$(\ref{eq40})$
may  be useful for investigating the Seifert conjecture in the context 
of naked singularities arising due to spherical collapses, which are 
being investigated (see  \cite{Sin98,Des98} and references therein).
$F=0$ in Eq.  $(\ref{eq40})$ gives one obtained  earlier (see in \cite{Vir97}).
For the JNW spacetime this equation immediately gives
\begin{equation}
E = M.
\label{eq44}
\end{equation}
The energy expression is independent of the radial distance and therefore
the entire energy is confined to the singularity. This is a counterexample
to the Seifert conjecture. However, for the purely scalar field ($M=0$ in the
JNW solution), the globally naked strong curvature singularity, $r=2q$, is
massless. This  supports the Seifert conjecture. The ``scalar charge''
in the JNW spacetime does not contribute to the  energy, because contributions
from  the matter and the field energy cancel. 
The anonymous referee mentioned that there are some indications that 
the JNW solution does not occur generically. Therefore, the
present counterexample to the Seifert conjecture may not be taken seriously.
However, if Eq. $(\ref{eq40})$ can be considered to be the  correct expression
for the energy distribution, then Eq. $(\ref{eq44})$
demonstrates that one cannot prove the Seifert conjecture in generality.

Now we summarize these in the following.
The possibility of the energy localization in general relativity has been
debated and there are mutually contradicting viewpoints on this 
issue\cite{MCB}. According to Bondi, a non-localizable form of energy is
inadmissible in relativity and its location can in principle be 
found\cite{MCB}. In fact, a  unanimously agreed precise definition of
local or even quasilocal mass   would have been very much useful to
understand some important issues in relativity. For instance, for 
investigating the Seifert conjecture (we have discussed in this paper)
and the {\em hoop conjecture}\cite{Tho72}, which states that  horizons form
when and only when a mass $M$ gets compacted into a region whose circumference
in {\em every direction} $\leq 2 \pi M$, these concepts are useful.
However, no adequate prescription is known and it is not clear if it is
possible at all. The results obtained using the energy-momentum complexes
are usually not taken seriously, because their uses are restricted to
``Cartesian coordinates'' and the quasilocal mass definitions are also
not satisfactory.
Bergqvist\cite{Ber92} showed that not any two of seven quasilocal mass
definitions he considered  give the same result for the Reissner-Nordstr\"{o}m
as well as Kerr spacetimes. The well-known Penrose quasilocal mass
definition could not deal with the Kerr metric\cite{BT94}.
On the other hand, several energy-momentum complexes
are known to give the same and ``reasonable'' result for many well-known
solutions.  We showed that different energy momentum complexes
disagree when they are evaluated in Schwarzschild Cartesian coordinates and 
give the same result in Kerr-Schild Cartesian coordinates;
however, the Einstein energy-momentum complex still gives consistent results in
both cases. It is not clear why different definitions ``coincide'' when
calculations are carried out in Kerr-Schild Cartesian coordinates, but
disagree in Schwarzschild Cartesian coordinates. At this stage it is not 
known if this is accidental or points out something interesting.
Any example or counterexample to the Seifert conjecture may not be taken 
seriously unless an adequate prescription for localization or 
quasilocalization  of mass  is known and is applied.

\acknowledgments
Thanks are due to  G. F. R. Ellis, Kerri (R.P.A.C.) Newman, and P. C. 
Vaidya for perusal of this manuscript and for valuable suggestions, 
and  to  J. M. Aguirregabiria, A. Chamorro, F. I. Cooperstock,  and K. P. Tod 
for helpful correspondence.  This research was supported by FRD, S. Africa.

\end{document}